\documentclass[final]{elsart3p}
\usepackage{graphicx}
\usepackage{subcaption}

\usepackage{amsmath}
\usepackage{amsfonts}
\usepackage{amssymb}
\usepackage{calrsfs}

\usepackage{multicol} 
\usepackage{cite}
\usepackage{appendix}
\usepackage{emptypage}
\usepackage{cancel}
\usepackage[font=small,labelfont=bf]{caption}
\usepackage{fancyvrb}
\usepackage{gensymb}
\usepackage{verbatim}
\usepackage{wrapfig}
\usepackage{units}
\usepackage{bm}
\usepackage{url}
\usepackage{xfrac}
\usepackage{xcolor}
\usepackage[english]{babel}
\usepackage{blindtext}
\usepackage{setspace}
\usepackage{csquotes}
\usepackage[a4paper,margin=2cm]{geometry}

\usepackage{array}
\newcolumntype{L}[1]{>{\raggedright\let\newline\\\arraybackslash\hspace{0pt}}m{#1}}
\newcolumntype{C}[1]{>{\centering\let\newline\\\arraybackslash\hspace{0pt}}m{#1}}
\newcolumntype{R}[1]{>{\raggedleft\let\newline\\\arraybackslash\hspace{0pt}}m{#1}}

\renewcommand{\v}[1]{\ensuremath{\mathbf{#1}}} 
\newcommand{\gv}[1]{\ensuremath{\mbox{\boldmath$ #1 $}}} 
\newcommand{\abs}[1]{\left| #1 \right|} 
\newcommand{\pd}[2]{\frac{\partial #1}{\partial #2}} 
 
\newcommand{\grad}[1]{\gv{\nabla} #1} 
\newcommand{\myparallel}{{\mkern3mu\vphantom{\perp}\vrule depth 0pt\mkern2mu\vrule depth 0pt\mkern3mu}}
\let\baraccent=\= 
\renewcommand{\=}[1]{\stackrel{#1}{=}} 

\journal{Computer Physics Communications}

\begin{document}

	\begin{frontmatter}
	\title{\textbf{A novel flexible field-aligned coordinate system for tokamak edge plasma simulation}}
	\date{\today}
	
	\author{J.Leddy\corauthref{cor}}
	\corauth[cor]{Corresponding author.}
	\ead{jarrod.leddy@york.ac.uk}
	\author{$^{1,2}$, B.Dudson$^1$, M.Romanelli$^2$, B.Shanahan$^1$, N.Walkden$^2$}
	\address{$^1$Department of Physics, University of York, Heslington, York YO10 5DD, UK}	
	\address{$^2$CCFE, Culham Science Centre, Abingdon, Oxfordshire OX14 3DB, UK}

    \begin{abstract}
      Tokamak plasmas are confined by a magnetic field that limits the particle and heat transport perpendicular to the field.  Parallel to the field the ionised particles can move freely, so to obtain confinement the field lines are ``closed" (ie. form closed surfaces of constant poloidal flux) in the core of a tokamak.  Towards, the edge, however, the field lines begin to intersect physical surfaces, leading to interaction between neutral and ionised particles, and the potential melting of the material surface.  Simulation of this interaction is important for predicting the performance and lifetime of future tokamak devices such as ITER.  Field-aligned coordinates are commonly used in the simulation of tokamak plasmas due to the geometry and magnetic topology of the system.  However, these coordinates are limited in the geometry they allow in the poloidal plane due to orthogonality requirements.  A novel 3D coordinate system is proposed herein that relaxes this constraint so that any arbitrary, smoothly varying geometry can be matched in the poloidal plane while maintaining a field-aligned coordinate.  This system is implemented in BOUT++ and tested for accuracy using the method of manufactured solutions.  A MAST edge cross-section is simulated using a fluid plasma model and the results show expected behaviour for density, temperature, and velocity.  Finally, simulations of an isolated divertor leg are conducted with and without neutrals to demonstrate the ion-neutral interaction near the divertor plate and the corresponding beneficial decrease in plasma temperature.
    \end{abstract}
    	
  \end{frontmatter}
  
\section{Introduction}
The plasma in the core of a tokamak is confined by magnetic fields that do not intersect any physical surfaces, but instead twist endlessly to form closed surfaces of poloidal flux.  The separatrix marks the dividing line between these ``closed" field lines and ones that are ``open" (ie. ones that intersect the material of the tokamak).  These open field lines are designed to intersect the divertor, which is made to withstand the particle and heat flux that escapes the core of today's machines.  In future devices such as ITER, however, the power flux will be too high ($>10$MW/m$^2$) for any known material to withstand for prolonged periods of time \cite{Pitts2011}.  By injecting neutral gas into the divertor region the plasma can be driven into a detached regime where the majority of the plasma power is radiated away before the plasma reaches the divertor, significantly lowering the plasma power deposited on it.  It is essential to accurately simulate such detached plasmas to predict the heat loads that will remain for these larger future devices \cite{Ricci2015}.  For this, it is important to match the simulation grid to the geometry of the divertor, which many codes currently do in the 2D poloidal plane, but the 3D extent remains unoptimised to the tokamak geometry due to the field-aligned nature of the plasma perturbations.
\\
\indent In tokamak plasmas, waves and instabilities are elongated along the magnetic field, while the perpendicular structures are small (on the order of the Larmor radius).  Therefore, when simulating an edge plasma it is desirable to also have a coordinate system and grid that are aligned along the field.  One can derive a set of coordinates related to standard orthogonal tokamak coordinates $(\psi, \theta, \phi)$ where one coordinate is aligned to the field.  Such a coordinate system allows for resolution along the field line to be sparser as is appropriate for the large structures, while maintaining fine resolution perpendicular to the magnetic field.  The typical method for doing this is to keep the radial flux coordinate $\psi$, but to replace the toroidal angle $\phi$ and the poloidal angle $\theta$ with a shifted toroidal angle $z$ and field-aligned coordinate $y$, respectively \cite{Dudson2009,Ribeiro2010}.  The mathematical derivation of this is detailed in the next section, but qualitatively this implies that if $\psi$ and $z$ are held constant while $y$ is increased, one will progress along the field line on a helical path around the torus.  The toroidal angle changes as one moves in $y$, implying that these coordinates are no longer orthogonal.
\\
\indent Though this system solves the problem of resolution, it leaves other problems un-addressed.  Namely, the grid is restricted in shape in the poloidal plane because the $\psi$ coordinate is orthogonal to the poloidal projection of $y$.  If this constraint is lifted by deriving a new set of coordinates that are both field-aligned but also non-orthogonal in $\psi$ and $y$, there is freedom to define a grid that matches the geometry of a specific tokamak in the divertor region.  This is especially useful for the simulation of neutrals because they do not follow the field, so a wall-conforming grid is necessary.  In this paper, a novel coordinate system that allows such freedom is presented, tested, and utilised for divertor plasma simulations.
\\
\indent An important distinction needs to be made between this new system and current coordinate systems in use in plasma edge codes such as SOLPS \cite{Schneider1992} and EDGE-2D \cite{Radford1996}.  These codes are 2D, so though they do allow non-orthogonality in the poloidal grid, they do not have a field-aligned coordinate.  The coordinate system derived in this paper allows for 3D plasma edge simulation grids to be defined with non-orthogonalities in the poloidal plane while also maintaining a field-aligned coordinate.

\subsection{Standard field-aligned coordinates}
\begin{figure}[h]
   	\centering
  	\includegraphics[width=\linewidth]{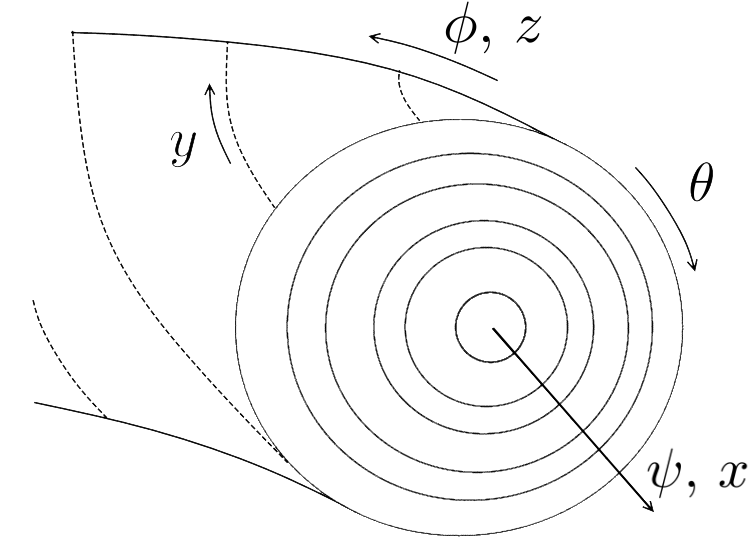} 
  	\caption[Geometry of field-aligned coordinate system]{The geometry described by the coordinate system posed in equations \ref{eqn:old_coords}.}
  	\label{fig:field-aligned coordinates}
\end{figure}

\indent In the derivation of these coordinates, standard symbols for tokamak geometry are utilised for the toroidal, poloidal, and radial flux coordinates - $\phi$, $\theta$, and $\psi$ respectively \cite{DHaeseleer1991}.  These coordinates form a right-handed, orthogonal coordinate system as shown in figure \ref{fig:field-aligned coordinates}.  The standard field-aligned coordinate system is defined as
\begin{equation}
	\begin{aligned}
		x &= \psi \vphantom{\frac11}\\
		y &= \theta \\
		z &= \phi - \int_{\theta_0}^{\theta}\nu \; d\theta
	\end{aligned} 
	\label{eqn:old_coords}
\end{equation}
where the local field line pitch is given by
\begin{equation}
\nu(\psi,\theta) = \pd{\phi}{\theta} = \frac{\v{B}\cdot \grad{\phi}}{\v{B}\cdot \grad{\theta}} = \frac{B_{\phi}h_{\theta}}{B_{\theta}R}.
\end{equation}
with toroidal field $B_{\phi}$, poloidal field $B_{\theta}$, major radius $R$, and poloidal arc-length $h_{\theta}$.
Figure \ref{fig:field-aligned coordinates} shows the geometry described by the coordinate system in equations \ref{eqn:old_coords}.  It is important to notice that the shift added to the $z$-coordinate causes the $y$-coordinate to be field-aligned.  The $x$-coordinate remains perpendicular to the poloidal projection of the $y$-coordinate, limiting choice of poloidal geometry.
\\
\indent The contravariant basis vectors are then found by taking the gradient of each coordinate, using $\nabla = \grad{\psi}\pd{}{\psi} + \grad{\theta}\pd{}{\theta} + \grad{\phi}\pd{}{\phi}$ to calculate
\begin{equation}
\begin{aligned}
\grad{x} &= \grad{\psi} \vphantom{\frac11}\\
\grad{y} &= \grad{\theta} \\
\grad{z} &= \grad{\phi} - \nu\grad{\theta} - I\grad{\psi}
\end{aligned}
\end{equation}
with
\begin{equation}
I =  \int_{\theta_0}^{\theta} \pd{\nu}{\psi} \; d\theta. \\
\end{equation}
The magnetic field can be written in Clebsh form \cite{Haeseleer1991},
\begin{equation}
\v{B} = \grad{x} \times \grad{z} = \frac{1}{\v{J}}\v{e}_y \label{eqn:Clebsh form}
\end{equation}
so the coordinate system is field aligned.  The contravariant and covariant metric tensors are defined as
\begin{equation}
\begin{aligned}
g^{ij} &= \grad{u}^i\cdot\grad{u}^j \\
g_{ij} &= \v{e}_i \cdot \v{e}_j \label{eqn:metric_tensor_definitions}
\end{aligned}
\end{equation}
where $\v{e}_{i} = \v{J}(\grad{u}^j \times \grad{u}^k)$ and $u^i$ indicates a particular coordinate.  Using the following identities
\begin{eqnarray}
\grad{\psi} = R\abs{B_{\theta}} \:\:\:\:\:\:&\:\:\:\:\: \grad{\theta} = \abs{h_{\theta}}^{-1} \:\:\:\:\:\:&\:\:\:\:\: \grad{\phi} = R^{-1} \label{eqn:metric physical identities}
\end{eqnarray}
the contravariant metric tensor can be rewritten as
\begin{eqnarray}
g^{ij} = 
\left[ {\begin{array}{ccc}
	(RB_{\theta})^2	&	0	&	-I(RB_{\theta})^2\\ \\
	\cdots	&	h_{\theta}^{-2} 	&	\nu h_{\theta}^{-2}\\ \\
	\cdots	&	\cdots	&	I^2(RB_{\theta})^2 + \nu^2h_{\theta}^{-2} + R^{-2}\\
	\end{array} }  
\right] \label{eqn:fa-contravariant}
\end{eqnarray}
To calculate the covariant metric tensor, one must first find the Jacobian of the system, which is given by
\begin{equation}
\v{J}^{-1} = \grad{x}\cdot (\grad{y}\times \grad{z}) \label{eqn:Jacobian definition}
\end{equation}
thus
\begin{equation}
\v{J} = \frac{h_{\theta}}{B_{\theta}}
\end{equation}
The covariant metric tensor, defined as $g_{ij}$ in equation \ref{eqn:metric_tensor_definitions}, is then calculated as
\begin{equation}
g_{ij} = 
\left[ {\begin{array}{ccc}
	I^2R^2 + (RB_{\theta})^{-2} & B_{\phi}h_{\theta}IRB_{\theta}^{-1} & IR^2\\ \\
	\cdots	&	h_{\theta}^2 + R^2\nu^2	&	\nu R^2\\ \\
	\cdots	&	\cdots	&	R^2\\
	\end{array} }  
\right] \label{eqn:fa-covariant}
\end{equation}
These co- and contravariant metric tensors can be used within simulations to perform operations, such as the parallel gradient, $\nabla_{\myparallel} = \hat{b}\cdot\v{\nabla} = \frac{1}{\vec{JB}}\pd{}{y}$, in the correct geometry \cite{Ribeiro2010}.

\section{Flexible field-aligned coordinates} \label{sec:new coords}
\begin{figure}[h]
   	\centering
  	\includegraphics[width=0.7\linewidth]{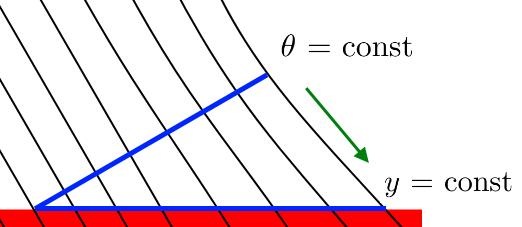} 
  	\caption[Geometry of $y$-shift term]{A physical picture of why the $y$-shift term (indicated by the green arrow) is needed and how lines of constant $\theta$ compare to lines of constant $y$.}
  	\label{fig:yshift}
\end{figure}

Near the divertor, this standard field-aligned system suffers from the inability to match the physical geometry of the divertor surface due to the orthogonality constraint in the poloidal direction.  Figure \ref{fig:yshift} shows a line of constant $\theta$, which represents a grid in the standard field-aligned coordinates.  It is ideal to shift this line so that it lies on the divertor plate, which requires a shift in the $\theta$ coordinate.  Though such a coordinate system is already utilised in many plasma codes for 2D simulations, a new set of coordinates is needed to allow a 3D simulation mesh to be aligned the divertor (or any smoothly varying) geometry in the poloidal plane while maintaining field-alignment, as well.  To derive these coordinates, the following system is defined by analogue to equation \ref{eqn:old_coords}:
\begin{equation}
\begin{aligned}
x &= \psi \vphantom{\frac11} \label{eqn:x-coordinate definition}\\
y &= \theta - y_{\text{shift}}  \\
z &= \phi - z_{\text{shift}}
\end{aligned}
\end{equation}
such that the shift in $y$ ($y_{\text{shift}}$) allows for the $x$-coordinate to be aligned with any arbitrary geometry in the poloidal plane.  Likewise the shift in $z$ ($z_{\text{shift}}$) enables the $y$-coordinate to follow an arbitrary geometry toroidally.  As is standard in field-aligned coordinates the $z_{\text{shift}}$ will be defined to ensure that the $y$-coordinate follows the magnetic field line, as demonstrated in the previous section.
\\
\indent For a coordinate system to uniquely define all points in space it must obey
\begin{equation}
\pd{x}{y} =  \pd{x}{z} = \pd{y}{x} =  \pd{y}{z} =  \pd{z}{x} =  \pd{z}{y} = 0.
\end{equation}
In this way one can derive the $y_{\text{shift}}$ by recognising that $\pd{y}{x} = 0$ so
\begin{equation}
0 = \pd{y}{x} = \pd{}{\psi}\left(\theta - y_{\text{shift}}\right) \;\; \rightarrow \;\; y_{\text{shift}} = \int_{\psi_0}^{\psi}\pd{\theta}{\psi}\;d\psi.
\end{equation} \\
\indent A non-orthogonality parameter (analogous to the field line pitch, but in the poloidal plane) is defined as $\eta = \pd{\theta}{\psi}$.  Similar to $\nu$, the field line pitch, $\eta$ is a function of $\psi$ and $\theta$.  This yields the final expression for the y-coordinate:
\begin{equation}
y = \theta - \int_{\psi_0}^{\psi}\eta\;d\psi. \label{eqn:y-coordinate definition}
\end{equation}\\
\indent Figure \ref{fig:yshift} demonstrates the physical functionality of the $y$-shift term in matching the divertor geometry.  The result of this shift, represented by the green arrow, is the alignment of the $x$-coordinate with the divertor plate.\\
\\
\indent The same method can be used to solve for $z_{\text{shift}}$ by this time recognising that $\pd{z}{y} = 0$,
\begin{equation}
0 = \pd{z}{y} = \pd{}{y}\left( \phi - z_{\text{shift}} \right) \;\;\;\;\; \rightarrow \;\;\;\;\;  z_{\text{shift}} = \int_{y_0}^{y}\pd{\phi}{y}\;dy
\end{equation}
however this needs further manipulation using equation \ref{eqn:y-coordinate definition} to obtain a final system dependent on established parameters.
\begin{equation}
\begin{aligned}
z_{\text{shift}} &= \int_{y_0}^{y}\pd{\phi}{y}\;dy \\
& = \int_{y_0}^{y}\pd{\phi}{\theta}\pd{\theta}{y}\;dy \\
&= \int_{y_0}^{y}\pd{\phi}{\theta}\left( 1 + \pd{}{y}\int_{\psi_0}^{\psi}\eta\;d\psi \right)\;dy
\end{aligned} 
\end{equation}
\indent As defined previously, the field line pitch $\nu=\pd{\phi}{\theta}$, yielding the final expression for the $z$-coordinate:
\begin{equation}
z = \phi - \int_{y_0}^{y}\nu\left( 1 + \pd{}{y}\int_{\psi_0}^{\psi}\eta\;d\psi \right)\;dy \label{eqn:z-coordinate definition}
\end{equation}\\
\indent With new coordinate definitions derived above and given in equations \ref{eqn:x-coordinate definition}, \ref{eqn:y-coordinate definition}, and \ref{eqn:z-coordinate definition}, the new more general covariant and contravariant metric tensors are then derived.  The contravariant basis vectors are found, as before, by taking the gradient of each coordinate.
\begin{equation}
\begin{aligned}
\grad{x} &= \grad{\psi} \vphantom{\frac11}\\
\grad{y} &= G\grad{\theta} - \eta\grad{\psi} \\
\grad{z} &= \grad{\phi} - H\grad{\theta} - I\grad{\psi} \label{eqn:contravariant basis vectors}
\end{aligned}
\end{equation}
where
\begin{equation}
\begin{aligned}
G &= \pd{y}{\theta} = 1-\pd{}{\theta}\int_{x_0}^{x}\eta dx \\
I &= \pd{z}{\psi} = \pd{}{\psi} \int_{y_0}^{y}\nu \left(1 + \pd{}{y}\int_{x_0}^{x}\eta\;dx \right)dy \\
H &= \pd{z}{\theta} = \pd{}{\theta} \int_{y_0}^{y}\nu \left(1 + \pd{}{y}\int_{x_0}^{x}\eta\;dx \right)dy.
\end{aligned}
\end{equation}\\
\indent These expressions cannot be simplified via the Leibniz integral rule, as was done previously, because $y$ is not independent of $\theta$ or $\psi$.  Since the magnetic field can still be written in Clebsh form \cite{DHaeseleer1991}, as in equation \ref{eqn:Clebsh form}, the system is still field aligned.  However, these changes now allow the system to match any smoothly varying geometry in the poloidal plane through choice of $\eta$.
\\
The metric tensors now have adjusted terms as well as an additional non-zero term ($g^{xy}$) reflecting the non-orthogonality of the $x$ and $y$ coordinates.
\begin{equation}
\begin{aligned}
g^{xx} &= (RB_{\theta})^2 \\
g^{yy} &= G^2 h_{\theta}^{-2} + \eta^2(RB_{\theta})^2 \\
g^{zz} &= I^2(RB_{\theta})^2 + H^2h_{\theta}^{-2} + R^{-2} \\
g^{xy} &= -\eta(RB_{\theta})^2 \\
g^{xz} &= -I(RB_{\theta})^2 \\
g^{yz} &= I\eta (RB_{\theta})^2 - GH h_{\theta}^{-2}
\end{aligned}
\end{equation}
The Jacobian of the system can still be calculated as before in \ref{eqn:Jacobian definition}, giving
\begin{equation}
J = \frac{h_{\theta}}{GB_{\theta}}
\end{equation}
Finally, the covariant metric tensor is calculated as
\begin{equation}
\begin{aligned}
g_{xx} &= (RB_{\theta})^{-2} + \left(\frac{h_{\theta}\eta}{G}\right)^2 + \left(\frac{RH\eta}{G} + IR\right)^2 \\
g_{yy} &= \frac{h_{\theta}^2}{G^2} + \frac{R^2H^2}{G^2} \\
g_{zz} &= R^2 \\
g_{xy} &= \frac{h_{\theta}^2\eta}{G^2} + \frac{R^2H}{G}\left( \frac{H\eta}{G} + I \right) \\
g_{xz} &= R^2\left(\frac{H\eta}{G}+I\right) \\
g_{yz} &= \frac{HR^2}{G}
\end{aligned}
\end{equation}
\\
\indent Importantly, in the limit where $x$ and the poloidal projections of all field lines are orthogonal (ie. the standard field-aligned system) $y=\theta$, so
\begin{equation}
\begin{aligned}
\eta&=0 \hspace{2.5cm} G=1 \\
H&=\nu \hspace{2.5cm} I= \displaystyle\int_{\theta_0}^{\theta}\pd{\nu}{\psi}d\theta
\end{aligned}
\end{equation} 
thus, the standard field-aligned metrics in equations \ref{eqn:fa-contravariant} and \ref{eqn:fa-covariant} are recovered.

\section[Testing the system]{Testing the Coordinate System and Metrics}
Ensuring that this substantial change to the simulation geometry has been implemented correctly requires the numerical accuracy of the system to be benchmarked.  The implementation and testing of this new coordinate system is done in BOUT++, an edge plasma simulation library developed by Ben Dudson, \emph{et al} \cite{Dudson2009}.  The numerical accuracy of the system is verified via the method of manufactured solutions (MMS) \cite{Salari2000}, which is a common method for testing the numerical validity of fluid simulations.  BOUT++ itself has been previously benchmarked using MMS to affirm that its numerical methods are accurate to the correct order \cite{Dudson2016}.  Any shortfalls in accuracy in this new test must then be due to the new coordinate system.

\subsection{Numerical accuracy}
To validate with MMS, a field $f(\psi,\theta,\phi,t)$ is defined and evolved using a simple advection model
\begin{equation}
\pd{f}{t} = \hat{Q}f + S
\end{equation}
where the operator $\hat{Q}=\pd{}{\psi}+\pd{}{\theta}+\pd{}{\phi}$ and $S(\psi,\theta,\phi,t)$ is a source term for the MMS.  An analytic function is chosen for $f=F$, and the source term is defined as
\begin{equation}
S = \pd{F}{t} - \hat{Q}F
\end{equation}
By doing this, we ensure that the numerical time derivative will be equal to the analytic time derivative in the case where the numerical $\hat{Q}$ is equivalent to the analytic $\hat{Q}$, since the source $S$ can be calculated analytically to machine precision.  In this way the numerical accuracy of the derivative operators is tested, as any error in them will propagate in time.  This has previously been done for BOUT++ showing that all second order methods are indeed accurate to second order \cite{Dudson2016}.  Verification of the new metric is then done by evaluating this equation on various non-orthogonal grids and observing the order of convergence as the grid spacing is changed.\\
\begin{figure}[h]
   \centering
    \begin{subfigure}{0.33\linewidth}
      \centering
      \includegraphics[width=\linewidth]{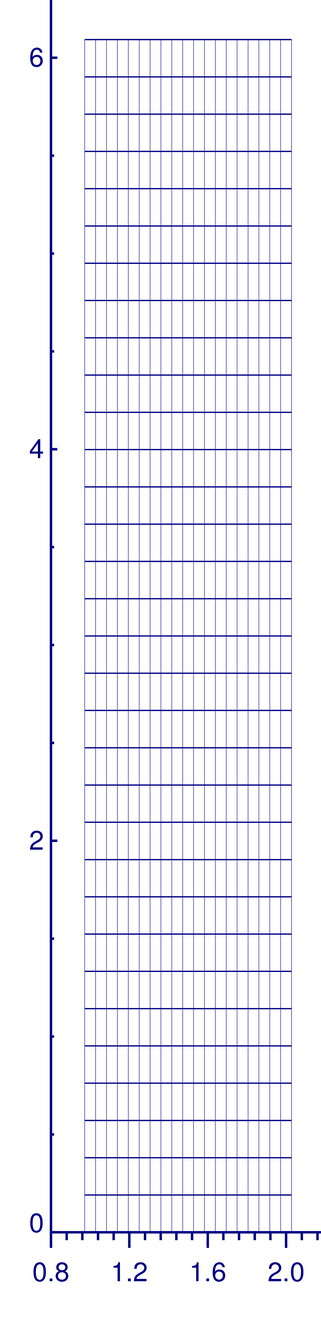} 
      \caption{}
      \label{fig:orthogonal_mesh1}
    \end{subfigure}%
    \begin{subfigure}{0.33\linewidth}
      \centering
      \includegraphics[width=0.925\linewidth]{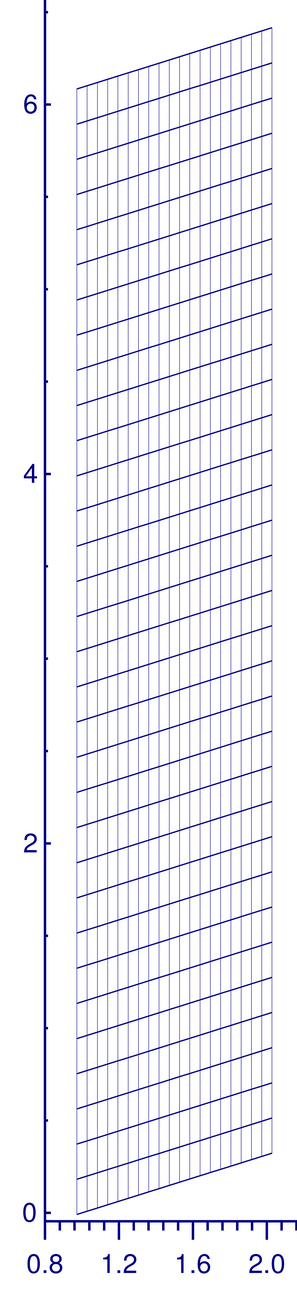}
      \caption{}
      \label{fig:nonorthogonal_mesh2}
    \end{subfigure}%
    \begin{subfigure}{0.33\linewidth}
      \centering
      \includegraphics[width=0.975\linewidth]{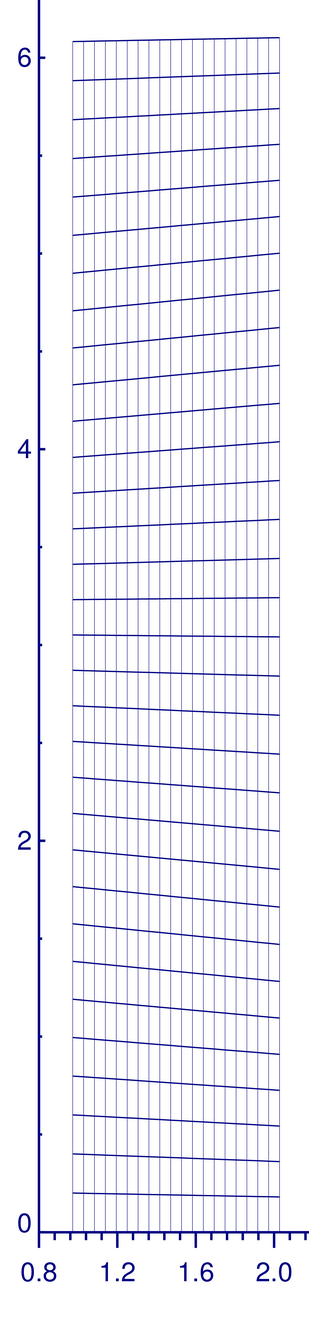}
      \caption{}
      \label{fig:varyingnonorthogonal_mesh}
    \end{subfigure}
    \caption[Meshes for metric verification]{The grids used for both the wave test case and the MMS verification.  (a) A fully orthogonal grid.  (b) A grid with constant non-orthogonality in the poloidal plane.  (c) A grid with sinusoidally varying grid spacing in the $y$-direction.}
    \label{fig:all_grids}
  \end{figure}\\
An analytic solution is defined
\begin{equation}
F(\psi,\theta,\phi,t) = \cos^2\left( \psi + \theta + \phi - t \right)
\end{equation}
which in turn provides the definition for the source term
\begin{equation}
S(\psi,\theta,\phi,t) = 8\sin(\psi+\theta+\phi-t)\cos(\psi+\theta+\phi-t).
\end{equation}\\
\indent This manufactured solution is chosen to satisfy the criteria laid out in \cite{Salari2000}.  Using a test grid and working from lowest to highest complexity in meshing, the new metric has been fully validated.  
\begin{table}[h]
	\begin{center}
		\caption[MMS convergence order for new metric]{Numerical scheme ordering as converged from 8x8x8 to 64x64x64 using the MMS.  Columns indicate non-orthogonality in the $x$-$y$ plane and rows describe non-orthogonality in the $y$-$z$ plane.}  \label{tab:nonorth}
		\begin{tabular}{ C{2cm}  C{1.9cm}  C{1.9cm}  C{2cm} } \hline
			& Orthogonal & Poloidal pitch & Poloidal shear \\
			& ($\eta=0$) & ($\eta=0.2$) & ($\eta=f(\psi,\theta)$) \\ \hline\hline
			No pitch ($\nu=0$) & 2.00  & 2.14  & 2.00  \\ 
			Const. pitch ($\nu=0.1$) & 2.02  & 2.04  & 2.02  \\ 
			Shear ($\nu=0.1x$) & 2.14  & 2.14 & 2.13  \\
			\hline
		\end{tabular}
	\end{center}
\end{table}\\
\indent The combinations of non-orthogonalities tested demonstrate at least 2\textsuperscript{nd} order convergence with Dirichlet boundary conditions.  Table \ref{tab:nonorth} shows the slope of the error norm, defined as the root mean squared error between $f$ and $F$ over the entire grid, as a function of grid spacing.  This indicates the order of convergence for the numerical methods in use in the simulation are second order as expected.  Similar convergence of at least 2\textsuperscript{nd} order is seen for the newly implemented Neumann boundary conditions, as well.  \\
\indent The $x-y$ grids used for these MMS verification tests are shown in figure \ref{fig:all_grids}.  Though the $z$-direction is not pictured, the 3-D location of the grid points is calculated through the pitch $\nu$ and non-orthogonality factor $\eta$ according to equations \ref{eqn:x-coordinate definition}, \ref{eqn:y-coordinate definition}, and \ref{eqn:z-coordinate definition}.  The field line pitches used for the test cases are $\nu=0$ for the orthogonal case, $\nu=0.1$ for the constant field line pitch case, and $\nu=0.1x$ for the magnetic shear case, with $x\in[0,1]$.  The $y$-grid spacing for the third grid (figure \ref{fig:all_grids}c) is defined by $y=\theta+b(0.5-x)\sin\theta$, where $\theta$ is equally spaced in $[0,2\pi]$, and $b$ determines the amount of non-orthogonality ($b=0.1$ for this case).  With this expression for $y$, there is no analytic form for $\eta$ so it was calculated numerically.  With the implementation proven numerically accurate, the following sections detail fluid simulations of plasma edge and divertors using the new coordinate system. 

\section{Application to divertor physics}
The X-point and divertor regions are the focus for simulations testing the effect of the new coordinate system for realistic geometries.  This is appropriate because these are regions where the flexible field-aligned coordinates are most pronounced (ie. where $\eta$ is the largest).  The first simulations involve a full tokamak poloidal cross-section and the second focus on an isolated divertor leg.\\
\indent In order to study physics problems in a realistic tokamak geometry, a grid generator called Hypnotoad \cite{Dudson2009} is used to create BOUT++ meshes from EFIT equilibria \cite{Lao1985}.  This generator was modified in order to create poloidally non-orthogonal meshes, and the calculations for the metrics are included in post-processing of the grids.
\begin{figure} 
	\centering
	\begin{subfigure}{0.5\linewidth}
		\centering
		\includegraphics[width=0.8\linewidth]{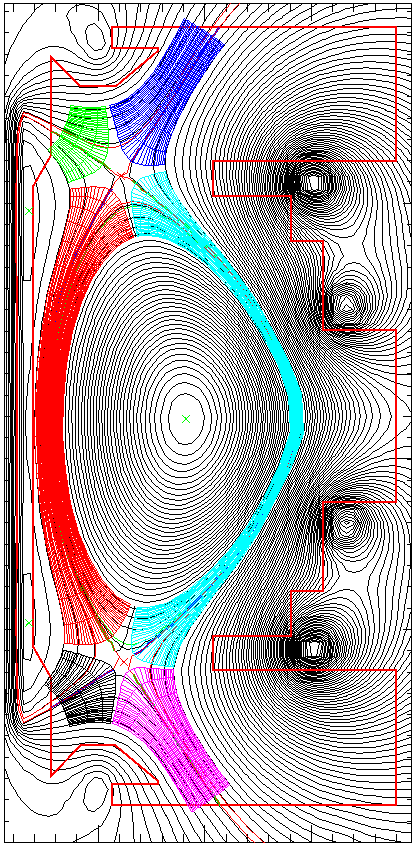} 
		\caption{}
		\label{fig:old_mesh}
	\end{subfigure}%
	\begin{subfigure}{0.5\linewidth}
		\centering
		\includegraphics[width=0.8\linewidth]{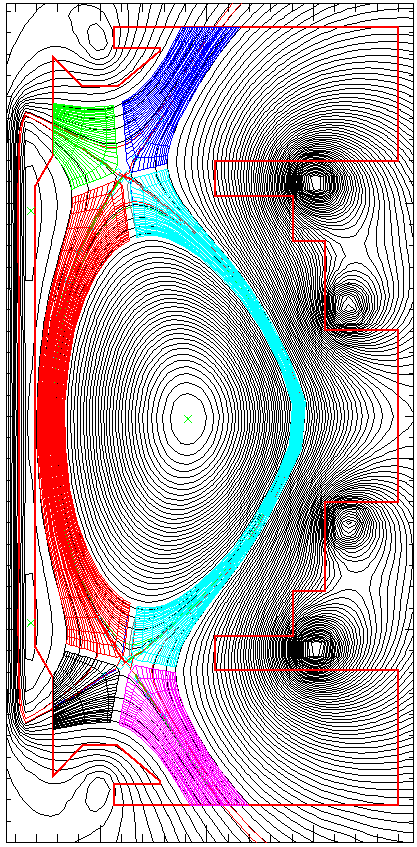}
		\caption{}
		\label{fig:new_mesh}
	\end{subfigure}
	\begin{subfigure}{\linewidth}
		\centering
		\includegraphics[width=0.5\linewidth]{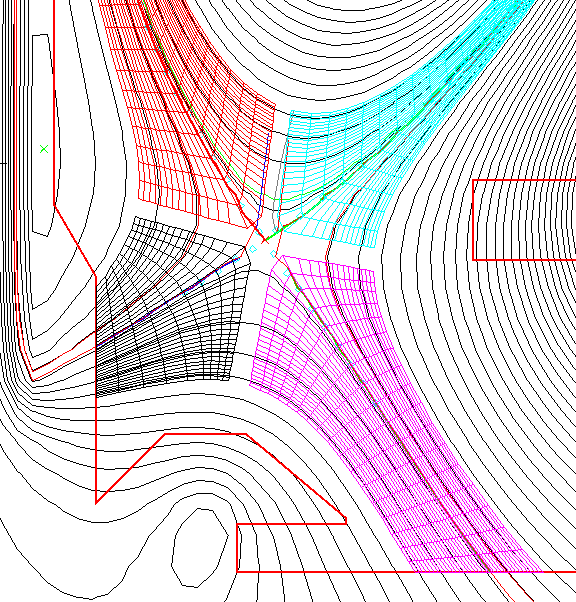}
		\caption{}
		\label{fig:xpt_mesh}
	\end{subfigure}
	\caption{(a) Orthogonal in the poloidal plane, this mesh cannot extend all the way to the divertor plate.  (b) This grid does extend to the divertor plate, but requires the new metric derived above. (c) The Cartesian nature of the X-point grid can be seen when the picture is zoomed for the new coordinate system case.  The slow change from orthogonal to non-orthogonal in the divertor leg as the plate is approached can also be seen.}\label{fig:mast_meshes}
\end{figure}

\subsection{Grid generation and processing} \label{sec:MAST}
Before the derivation of the new coordinate system, BOUT++ used the standard field-aligned coordinate system (equation \ref{eqn:old_coords}) requiring simulation meshes like that seen in figure \ref{fig:old_mesh}.  However, the mesh can now be constructed to match the geometry of the divertor as shown in figure \ref{fig:new_mesh}, allowing for more accurate simulations of the physics in this region.
\begin{figure}[h]
   	\centering
  	\includegraphics[width=0.7\linewidth]{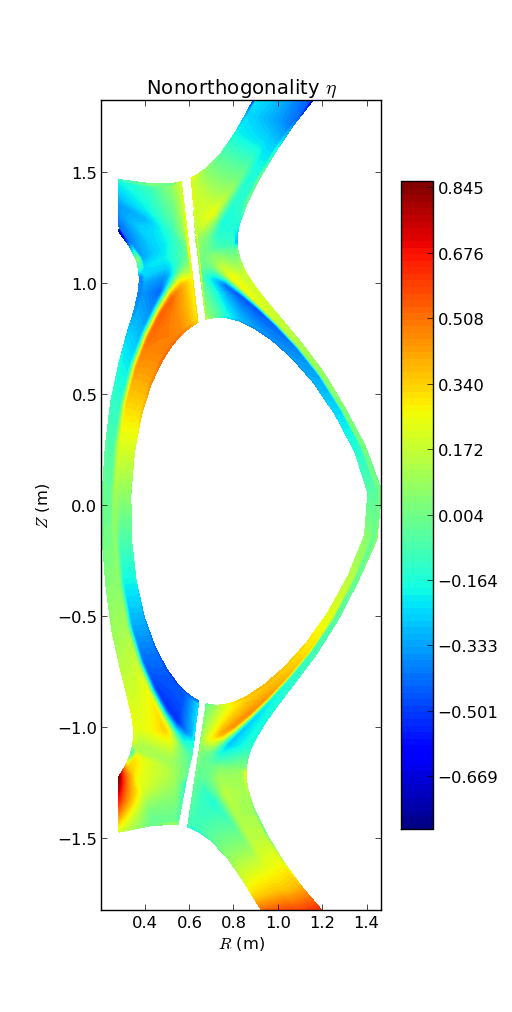} 
  	\caption[Contour of the non-orthogonality factor, $\eta$ on the new MAST mesh]{The non-orthogonality factor, $\eta$, is contoured onto the MAST mesh to highlight the differences between this mesh and the old one around the divertors and X-points.}
  	\label{fig:eta_contour}
\end{figure}
\\
\indent There are significant differences in these two grids at the divertor leg and also at the X-points.  The changes to the X-points are a benefit of this new coordinate system because it allows a more regular distribution of grid spacing in $y$ in this region, which increases the stability of simulations taking them further from the Courant-Freidrichs-Lewy condition limit \cite{Courant1967} and increasing the speed of the simulations.  It also allows data to be simulated closer to the X-point, though there is still a coordinate system singularity on the X-point itself as is common among all field-aligned systems.  The non-orthogonality of the new mesh is captured in the value $\eta$, which can be seen contoured in figure \ref{fig:eta_contour}.  It was not as straight-forward to calculate $\eta$ for these realistic meshes as it was for the meshes in figure \ref{fig:all_grids} created for the MMS verification.  In the generation of those grids, $\theta$ was defined and $y$ was derived from the by setting the value of $y_{\text{shift}}$ analytically and utilising equation \ref{eqn:y-coordinate definition}.  This process, however, is not possible for the realistic tokamak geometries, so $\eta$ needs to be calculated from the layout of the grid as produced by the grid generator.  This is done by realising that the non-orthogonality factor $\eta = \sin{\beta}$ where $\sfrac{\pi}{2}-\beta$ is the local angle between $y$ and $x$.  In this way, Hypnotoad was modified to include the calculation of $\beta$ and $\eta$ for each grid point, allowing for the calculation of the full metric tensors.

\subsection{Simulations in BOUT++}
Herein the standard field-aligned coordinates (equation \ref{eqn:old_coords}) are referred to as `orthogonal' and the flexible field-aligned coordinates (equations \ref{eqn:x-coordinate definition}, \ref{eqn:y-coordinate definition}, and \ref{eqn:z-coordinate definition}) are called `non-orthogonal' due to their projected behaviours in the poloidal plane.

\subsubsection{Plasma 2-D transport model}\label{sec:simcat}
There are many 3-dimensional turbulence models that have been derived for simulation of tokamak plasmas \cite{Yoshizawa2001}.  Since the flexible field-aligned coordinate system is ideal for simulating the edge and divertor, which is very collisional due to the low temperatures and densities, one might consider a drift-reduced model which has been shown to be valid for the edge \cite{Leddy2015}.  A model developed by Simakov and Catto \cite{Simakov2004}, cast into divergence-free form, and simplified to the fluid limit (ie. no current) is utilised for 2-D divertor and edge simulations in the new coordinate system using BOUT++.\\
\indent The Simakov-Catto model can be simplified to include only perpendicular and parallel transport via diffusion, conduction, and convection.  This is a useful tool for initialising a full turbulence run and also to check the basic behaviour of a system.  The reduced equations are\\
\begin{equation}
\begin{aligned}
\pd{n}{t} &= - \nabla_\myparallel \left(nv_{i\myparallel}\right) + \vec{\nabla}_\perp \cdot \left( D_n\nabla n \right) + S_n \\
\pd{(nv_{i\myparallel})}{t} &= - \nabla_\myparallel\left( nv_{i\myparallel}^2 \right) - \nabla_\myparallel p_e \\
\frac{3}{2}\pd{p_e}{t} &= - \frac{5}{2}\nabla_\myparallel\left( p_ev_{i\myparallel} \right) + v_i\nabla_{\myparallel}p_e + \vec{\nabla}\cdot\left(\kappa \nabla_\myparallel T_e\right) \\ 
&+ \frac{3}{2}\vec{\nabla}_\perp \cdot \left(D_n \nabla p_e \right) + S_p
\end{aligned}
\end{equation}
This system is simulated in two-dimensions (x-y) to evolve the flows within the system, but turbulence and electric effects (such as $\vec{E}\times\vec{B}$ drifts) are absent.  Sheath conditions are used for the boundary in contact with the divertor plate according to the constraints given by Loizu \cite{Loizu2012}.  The ion velocity at the plate is assumed to be the sound speed, as is the Bohm criterion, and this is then related to the electron velocity.
\begin{equation}
v_i = c_s = \sqrt{T_e}
\end{equation}
The electron temperature is assumed to have zero gradient at the divertor allowing a flow of heat to the plates.
\begin{equation}
\nabla_{\myparallel}T_e = 0
\end{equation}
The density and velocity gradients are set by assuming the ion flux has a zero parallel gradient.
\begin{equation}
\nabla_{\myparallel}n = -\frac{n}{c_s} \nabla_{\myparallel}v_i
\end{equation}
Finally, the pressure gradient is set by assuming zero temperature gradient and $p_e=nT_e$.
\begin{equation}
\nabla_{\myparallel}p_e = T_e \nabla_{\myparallel} n 
\end{equation}

\subsubsection{Tokamak simulation}
\begin{figure}
	\centering
	\includegraphics[width=0.9\linewidth]{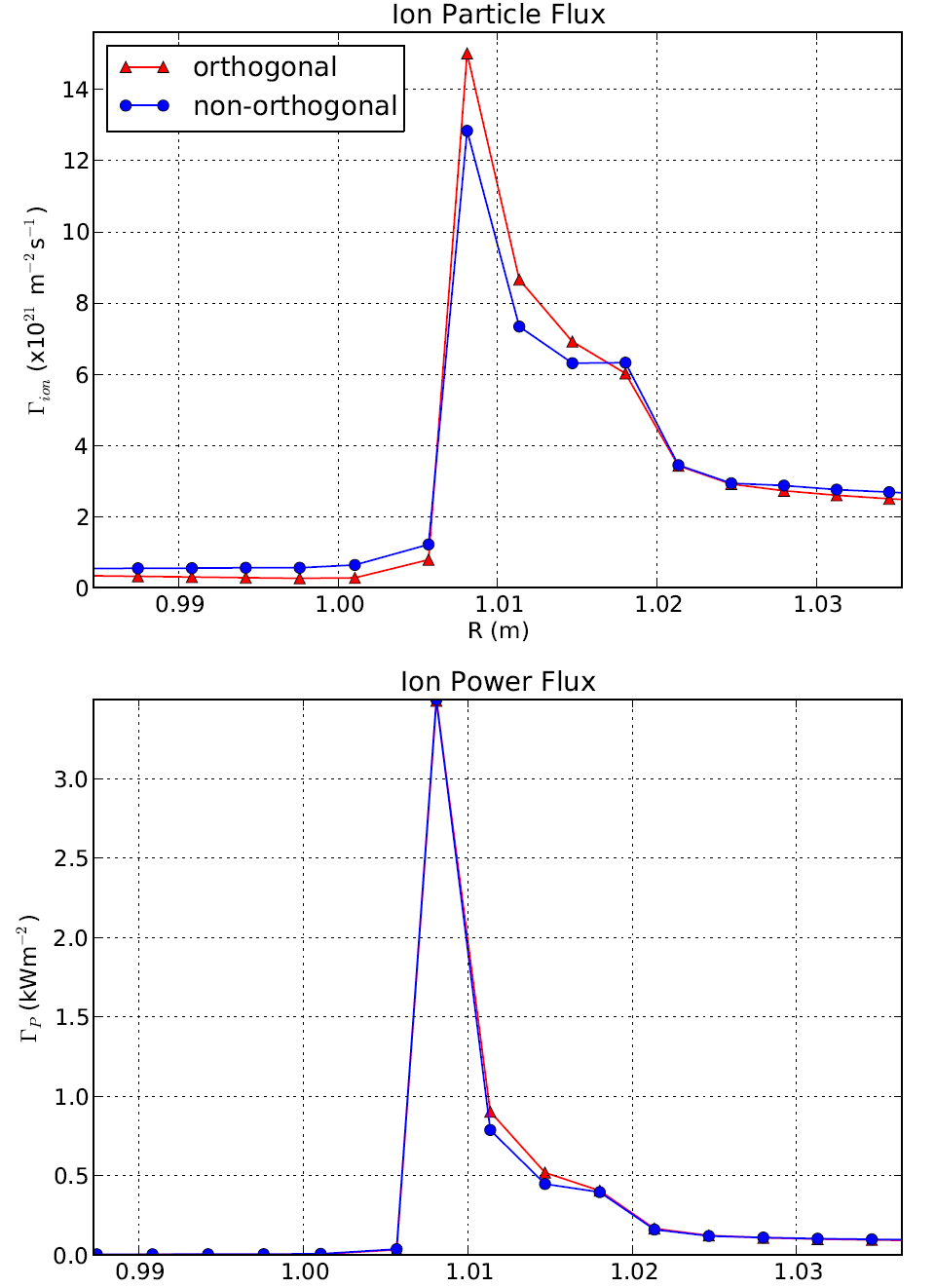}
	\caption[Plasma ion and power flux profiles at lower outer divertor strike point]{The particle (left) and power (right) fluxes are shown for both the orthogonal (red) and non-orthogonal (blue) grids.}
	\label{fig:fluxes_at_strikepoints}
\end{figure}

A full MAST lower double-null equilibrium was constructed, as in section \ref{sec:MAST}, to examine the behaviour of the plasma in the divertor region and around the X-point.  An L-mode plasma is simulated with the pedestal temperature at the inner boundary set to 295eV and the edge to 10eV, the core density to $10^{19}$m$^{-3}$ and edge density to $10^{18}$m$^{-3}$.  Using the grids shown in figure \ref{fig:mast_meshes} and the fluid model described in section \ref{sec:simcat}, a steady state is reached after $1.25$ms.  \\
\begin{figure}
	\centering
	\includegraphics[width=0.8\linewidth]{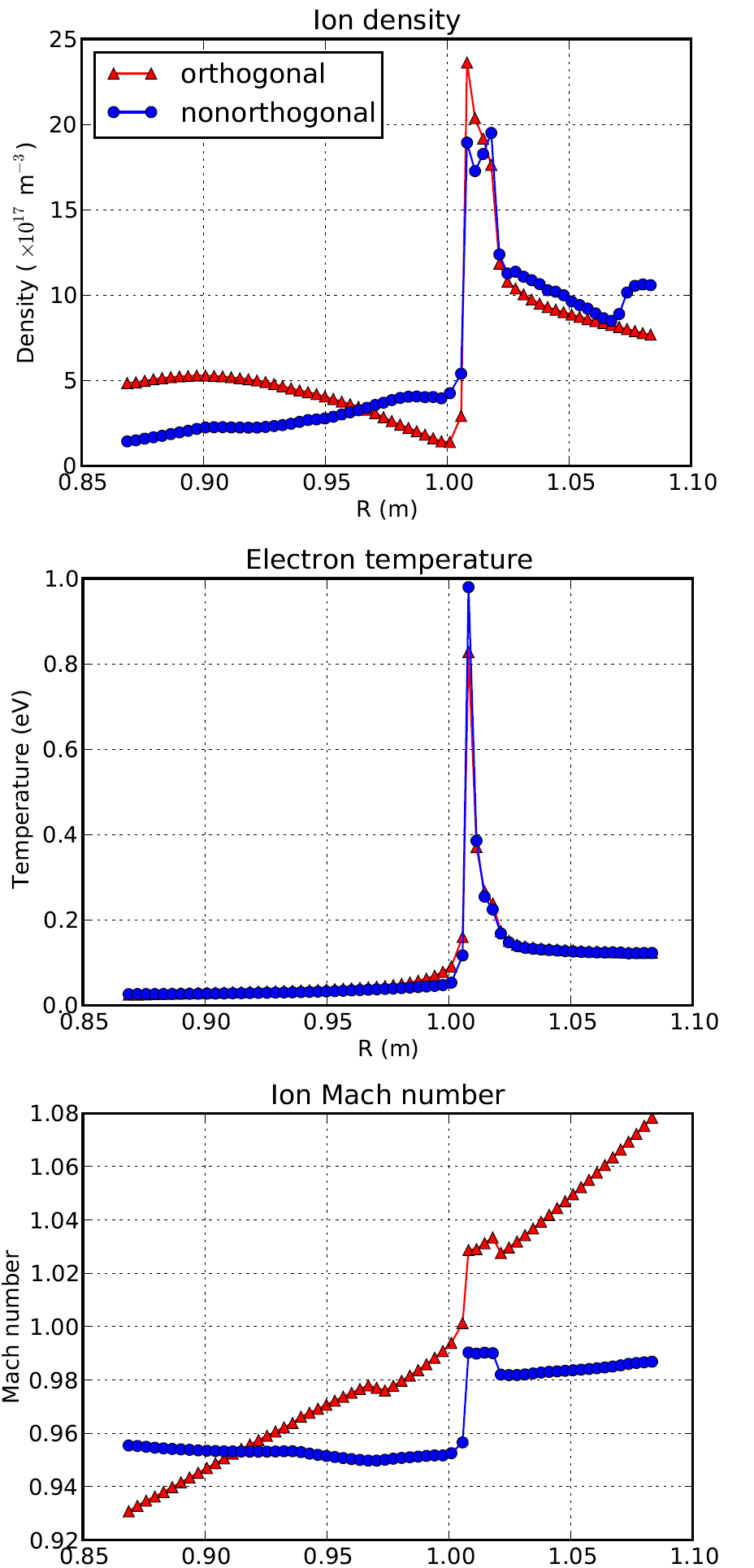} 
	\caption[Plasma field profiles at lower outer divertor strike point]{The density, temperature, and Mach number profiles at the divertor plate are shown for both the orthogonal (red) and non-orthogonal (blue) grids.}
	\label{fig:fields_at_strikepoints}
\end{figure}
\indent The Bohm boundary conditions set the Mach number to 1 on the cell boundary, which is halfway between the last grid cells and the first boundary cells in the $y$-direction.  For the non-orthogonal case this corresponds to the divertor plate, but for the orthogonal grid, this lies in an arbitrary distance form the plate, so values must be extrapolated and scaled based on the flux expansion, $\sfrac{B_{\phi}}{B_{\theta}}$, to obtain the quantitative results at the divertor plate.  Figure \ref{fig:fields_at_strikepoints} shows the density, and Mach number at the strike point for both the orthogonal and the non-orthogonal grids.  The densities and temperatures are very similar for both cases, while the Mach number shows a background linear profile for the orthogonal grid which is due the boundary conditions being set at the last grid cells, but the grid itself being at an angle to the plate.  This means that the boundary conditions are being places at a location inside the plasma instead of at the true boundary of the machine, a problem solved by the flexible field-aligned coordinate system. \\
\indent The density drops at the plate to a value around $2\times10^{18}$m$^{-3}$ and the temperature to $1$eV.  If neutral interactions were part of this simulation of the divertor region, plasmas of this temperature would be dominated by charge exchange reactions and detachment would occur.  Since there are no neutrals, however, the low temperature and density lead to small ion and power fluxes, shown in figure \ref{fig:fluxes_at_strikepoints}.  Both the particle and power flux are similar for the orthogonal and non-orthogonal case, as expected since the input fluxes are the same.  This affirms the validity of the implementation of the new coordinate system in BOUT++ and allows for more detailed and interesting simulations to be run, such as those including neutrals in the next section.

\subsubsection{Fluid neutral model} \label{sec:neutrals}
\begin{figure}[h]
	\centering
	\includegraphics[width=\linewidth]{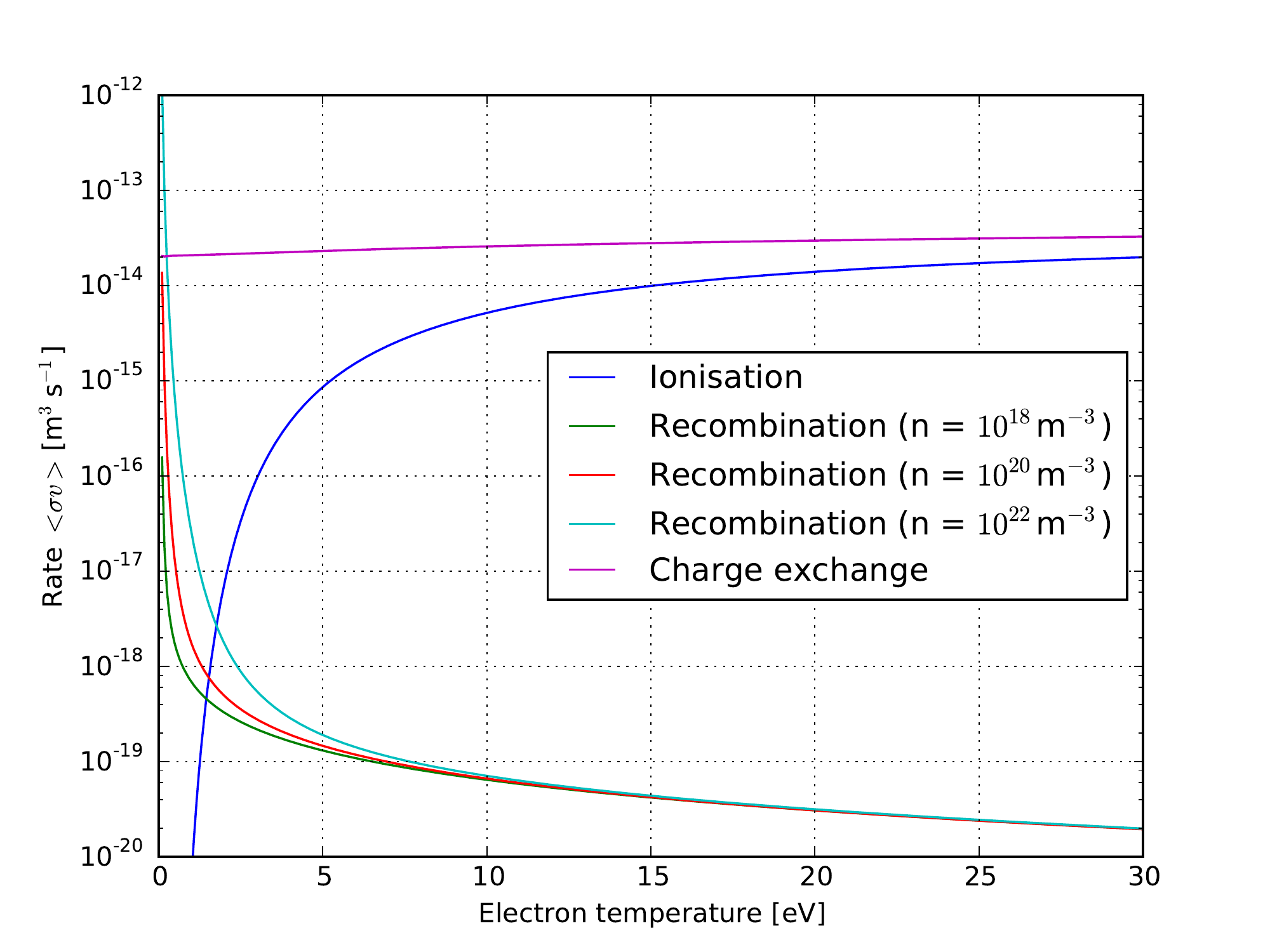} 
	\caption[Cross-sections for atomic processes]{The cross-section rates for ionisation, recombination, and charge exchange are pre-calculated for hydrogen species as a function of the plasma temperature (thanks to H. Willett).}
	\label{fig:atomic processes}
\end{figure}

The heat load from the plasma onto the divertor plates is a limiting factor in divertor design and tokamak operation.  In current devices, the heat load is acceptable, but for future fusion power plants, the predicted heat load would cause the divertor plate to melt \cite{Pitts2011}.  One method to solve this problem is to operate the tokamak in a detached regime, where neutral density rises at the plate allowing the plasma to radiate much of its energy in all angles instead of depositing it in a small area on the divertor plate.  The ease of entering this detached regime is dependent on the geometry of the divertor \cite{Loarte2001}, making this an ideal test case for the flexible field-aligned coordinates.  \\
\indent To simulate detachment, a model for the neutral behaviour as well as the neutral-plasma interaction is required.  The four atomic processes included to describe this interaction are ionisation, recombination, charge exchange, and radiation.  Ionisation, recombination, and charge exchange provide sources and sinks for density, energy, and momentum, each of which can be described by the following density rate coefficients (m$^{-3}$s$^{-1}$)
\begin{equation}
\begin{aligned}
\mathcal{R}_{iz} &= nn_n\left<\sigma v\right>_{iz} \qquad \mbox{\textrm{(Ionisation)}} \\
\mathcal{R}_{rc} &= n^2\left<\sigma v\right>_{rc}   \qquad \mbox{\textrm{(Recombination)}} \\
\mathcal{R}_{cx} &= nn_n\left<\sigma v\right>_{cx} \qquad \mbox{\textrm{(Charge exchange)}}
\end{aligned}
\end{equation}
where $n$ is the plasma density, $n_n$ is the neutral density, and $\left<\sigma v\right>$ is the cross-section (m$^3$s$^{-1}$) for the relevant process which is a function of the plasma temperature.  These cross-sections (shown in figure \ref{fig:atomic processes}) are pre-calculated and interpolated from a look-up table within the code.
Ionisation increases the plasma density while recombination decreases it, so the resulting density source is described as the difference:
\begin{equation}
S_n = \mathcal{R}_{iz} - \mathcal{R}_{rc}
\end{equation}
Recombination and charge exchange both remove momentum from the ions transferring it to the neutrals.  Therefore the sink of momentum is given by
\begin{equation}
F = - m_i\left[v_{\myparallel}\mathcal{R}_{rc} + \left( v_{\myparallel}-v_{n\myparallel} \right)\mathcal{R}_{cx} - v_{n\myparallel}\mathcal{R}_{iz}\right]
\end{equation}
where $v_{\myparallel}$ is the parallel ion velocity and $F$ can be described as a friction-like term.  Energy is transferred between the ions due to all three plasma-neutral interactive processes.  Ionisation provides an energy source to the plasma, while recombination removes energy from the plasma.  Charge exchange can technically act as a source or a sink for plasma energy depending on the relative temperature difference between the plasma and neutrals; however, it is unlikely for the neutrals to be hotter than the plasma, so in most cases charge exchange acts as a sink for plasma energy
\begin{equation}
E = \frac{3}{2}T_n\mathcal{R}_{iz} - \frac{3}{2}T_e\mathcal{R}_{rc} - \frac{3}{2}\left( T_e - T_n \right)\mathcal{R}_{cx}
\end{equation}
where $T_e$ is the plasma temperature and $T_n$ is the neutral temperature.  The plasma energy is also affected by radiation, which causes a loss of energy through photon emission and 3-body recombination, which heats the plasma at temperatures less than 5.25eV \cite{Stangeby2000}.  These two processes are calculated with
\begin{equation}
R = \left( 13.6\textrm{eV} - 1.09T_e \right)\mathcal{R}_{rc} - E_{iz}\mathcal{R}_{iz}
\end{equation}
where $E_{iz}=30$eV is the ionisation energy given by Togo \cite{Togo2013}.\\
\indent In order to calculate the interactions described above, the neutral density and temperature must be known.  There are various approximations that can be made for the behaviour of neutrals; however, for these simulations the neutrals have been included rigorously by co-evolving neutral density, pressure, and velocity with a standard system of fluid equations.
\begin{equation}
\begin{aligned}
\frac{\partial n_n}{\partial t} &= -\nabla\cdot\left(n_n\mathbf{v}_n\right) \\
\frac{\partial \mathbf{v}_n}{\partial t} &= - \mathbf{v}_n\cdot\nabla\mathbf{v}_n -\frac{1}{n_n}\nabla p_n + \frac{1}{n_n}\nabla\cdot\left(\mu \nabla\mathbf{v}\right)\\
 &+ \frac{1}{n_n}\nabla\left[ \left(\frac{1}{3}\mu + \zeta\right)\nabla\cdot\mathbf{v}_n\right] \\
\frac{\partial p_n}{\partial t} &= -\nabla\cdot\left(p_n\mathbf{v}_n\right) - \left(\gamma - 1\right)p_n\nabla\cdot\mathbf{v}_n \\
&+ \nabla\cdot\left(\kappa_n \nabla T_n\right)
\end{aligned}
\end{equation}
where $\mu$, $\zeta$, and $\kappa$ are constants describing the dynamic viscosity, bulk viscosity, and thermal conduction respectively.  For numerical stability, the neutral velocity is shifted to cylindrical coordinates, calculated, and then shifted back into the field-aligned coordinates.  This proves more stable and accurate since the neutrals are unaffected by the field lines and it removes derivatives of the metric tensor terms in the advection of vectors.  
\begin{figure}[h]
	\centering
	\includegraphics[width=0.9\linewidth]{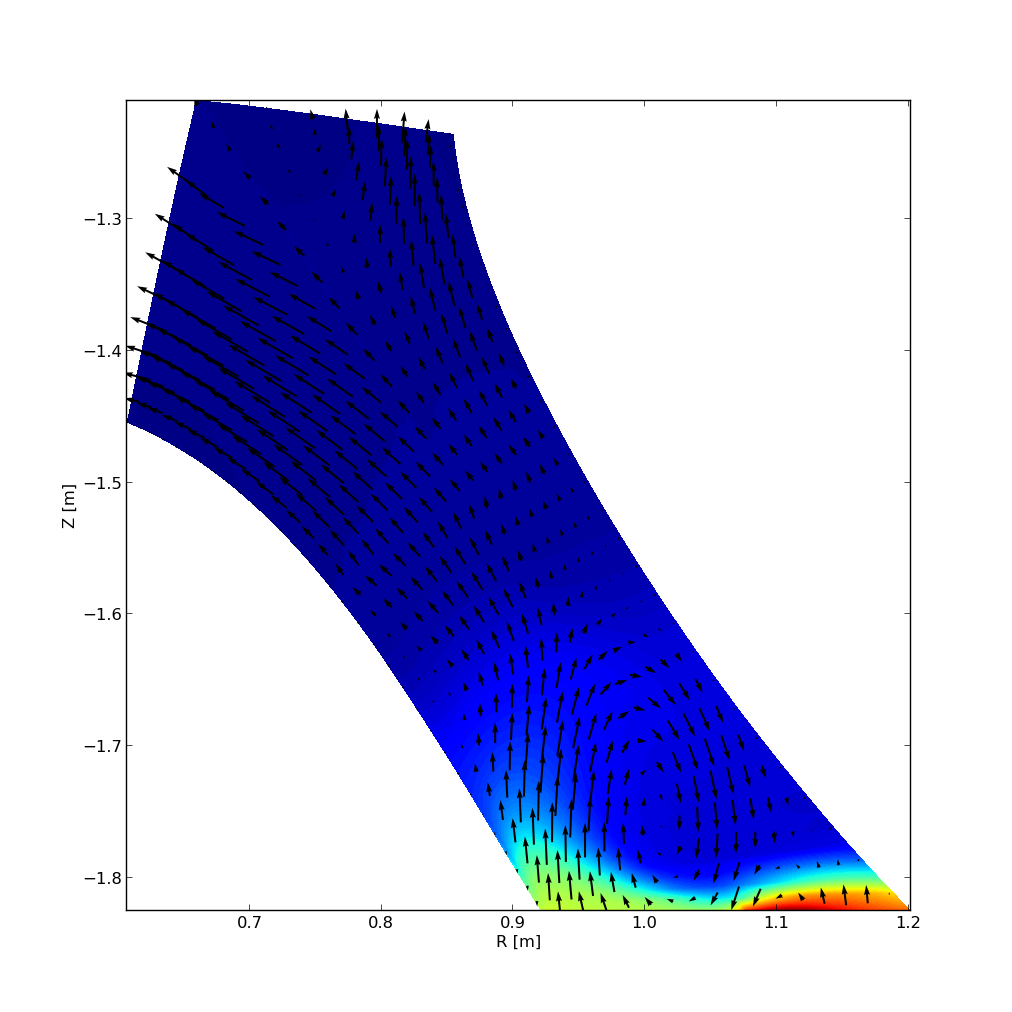} 
	\caption[Neutral density and flow in the divertor leg]{The coloured contours show the neutral density and the vectors indicate the flow direction and speed.  Neutrals flow away from the point of generation (ie. where the high plasma flux hits the plate) and cycle around the divertor leg.  The grid used for this simulation is described in section \ref{sec:iso_divertor_leg} and figure \ref{fig:divertor_leg_grid}.}
	\label{fig:neutral_flow_and_density}
\end{figure}
Figure \ref{fig:neutral_flow_and_density} shows neutrals that are generated at the divertor plate due to the plasma flux.  These neutrals then stream away from the plasma along the plate and then up the divertor leg.  The flow is shown to be cyclic as the neutrals make their way up the leg, back into the plasma where they are accelerated back to the plate.  The boundary conditions at the plate and sides of the legs are reflecting for neutrals, but the top of the leg allows outflow.

\subsubsection{Isolated divertor leg} \label{sec:iso_divertor_leg}
\begin{figure}[h]
   	\centering
  	\includegraphics[width=0.8\linewidth]{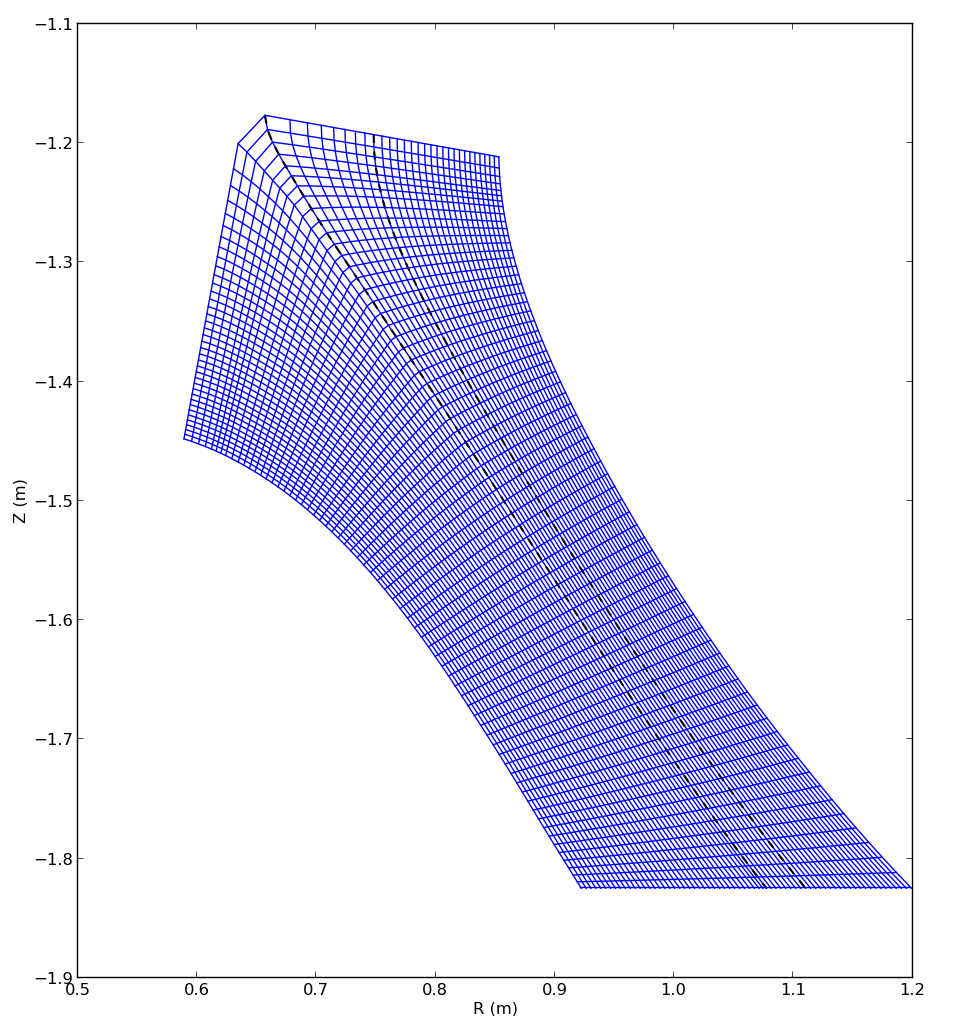} 
  	\caption{A simulation grid for an isolated outer, lower divertor leg allows for concentrated computational power on this region of interest.  The dotted black lines show the locations of the flux surfaces formed from the upper and lower X-point.}
  	\label{fig:divertor_leg_grid}
\end{figure}

\begin{figure}[h]
	\centering
	\includegraphics[width=\linewidth]{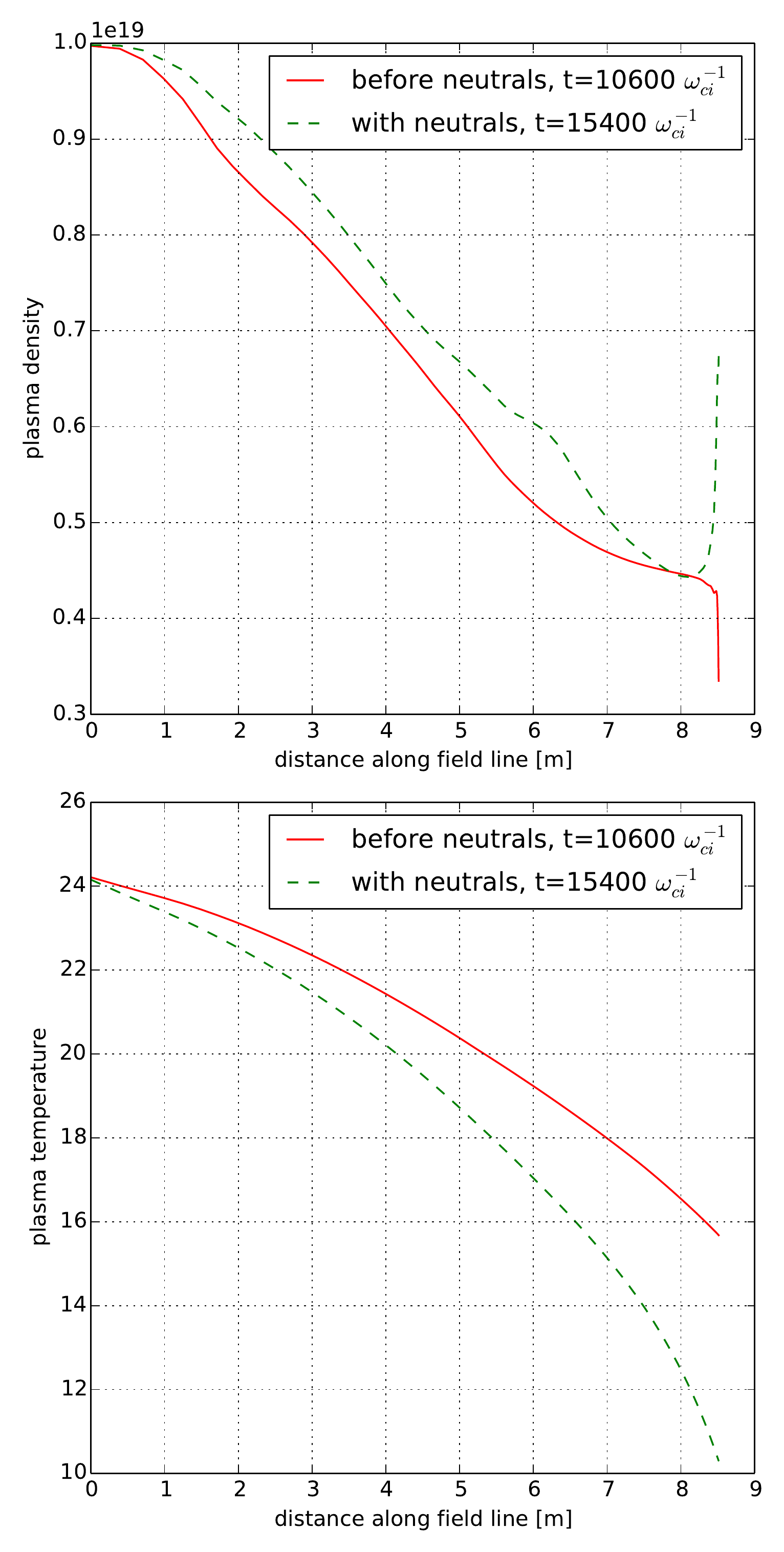}
	
	\caption[Effect of neutrals on parallel density and temperature plasma profiles]{Density and temperature are shown to decrease towards the divertor plate when no neutrals are present.  Once neutrals are introduced, the plasma density goes up near the plate due to a recycling source, while the temperature drops due to the energy lost through radiation.}
	\label{fig:parallel_profile_withneutrals}
\end{figure}

\begin{figure}[h]
	\centering
	\includegraphics[width=\linewidth]{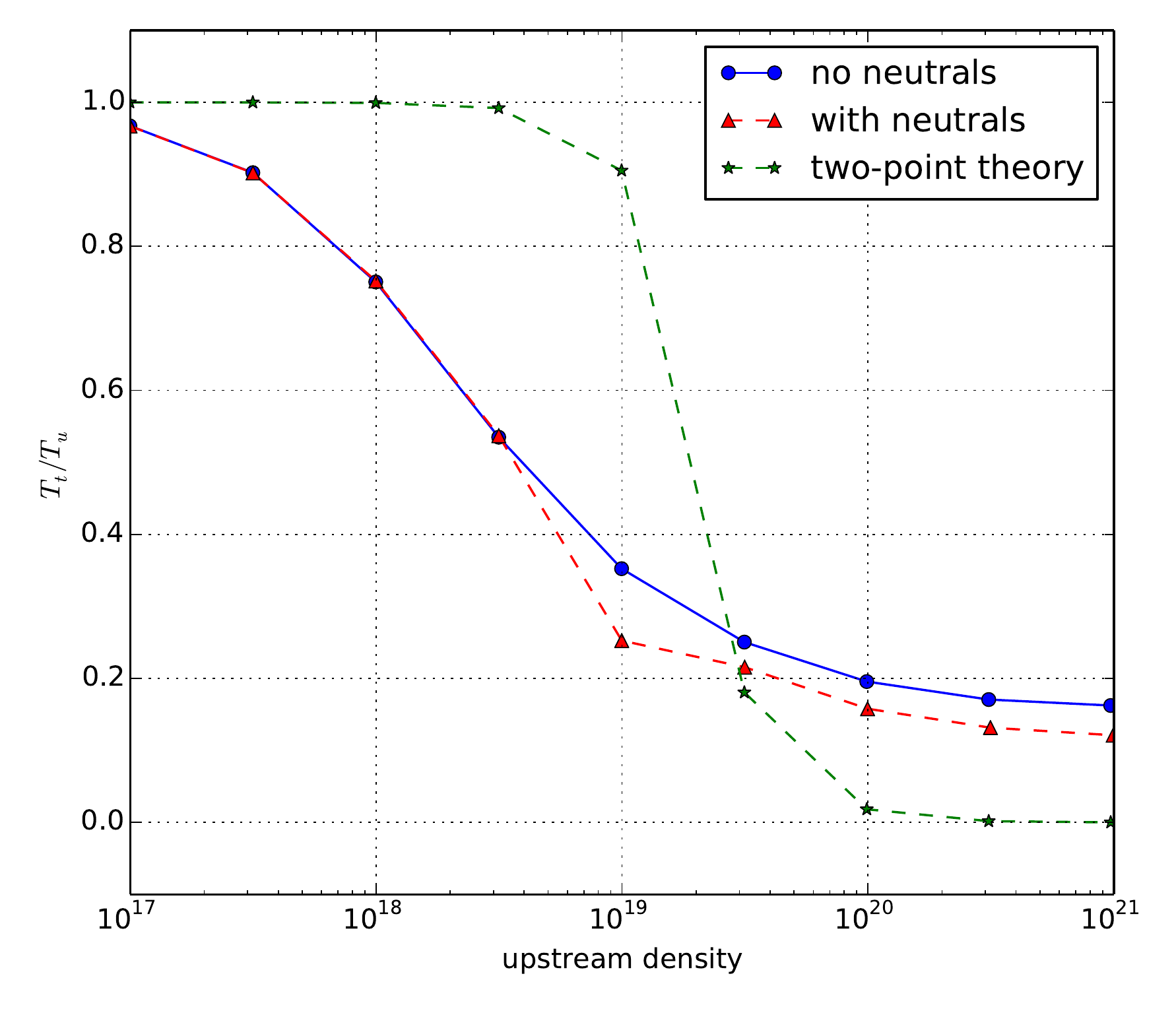} 
	\caption[The effect of the neutrals on $T_t/T_u$ as a function of upstream density]{The temperature decrease along the field line due to upstream density is seen to be more significant when neutrals are present.  The analytic solution to the two-point model is shown for comparison.}
	\label{fig:two-point-scan-detachment}
\end{figure}

In order to isolate the divertor leg, where the non-orthogonality is most pronounced, a grid is produced by amputating the leg from a full diverted plasma grid.  For this section the grid is a MAST lower, outer leg as shown in figure \ref{fig:divertor_leg_grid}.  For this to be sufficient an approximation for the core density and power fluxes must be made at the top of the leg.  This is accomplished by holding the density and temperature constant at the upper boundary in the outer SOL, while allowing the density and temperature to float with zero gradient boundary conditions in the private flux region.\\
\\
\indent Experimentally, detachment is seen to occur when the upstream density is high enough for the recycling at the divertor plate to cool the plasma below about 5eV \cite{Lipschultz1995}.  At this point, the plasma rapidly cools and recombines, forming a cloud of neutrals which radiates the heat away before the plasma reaches the divertor plate.  To see this in simulation, first a plasma fluid model (with no currents or neutrals) was run until equilibrium was reached.  The behaviour of the density and temperature should be \emph{similar} to the two-point model \cite{Stangeby2000}, which is a simplified view of the plasma behaviour that assumes parallel pressure conservation in a flux tube geometry with no flux expansion, yet not identical since the simulation has perpendicular diffusion and flux expansion.  \\
\indent Once steady state is reached in the fluid model, neutrals are added to the simulation and evolved using the fluid equations described in the previous section.  The recycling fraction is set to 95\%, and the expectation is that the $T_t/T_u$ should move to lower values at higher upstream densities, indicating a detached regime as the neutrals remove energy and momentum from the plasma.  Figure \ref{fig:parallel_profile_withneutrals} shows the parallel profiles of density and temperature before and after neutrals are added, and a clear drop in temperature is seen as well as a rise in plasma density near the plate due to ionisation.\\
\\
\indent Figure \ref{fig:two-point-scan-detachment} shows how the introduction of neutrals affects the $T_t/T_u$ curve - clearly, the neutrals are cooling and slowing the plasma through collisions and atomic processes.  Simulations with upstream density over $10^{19}$m$^{-3}$ were unable to complete due to numerical instability.  The mean-free path of the neutrals decreases as a function of density, so with higher density, the resolution requirement is significant near the plate where the neutrals are initially formed.  This problem is not due to the coordinate system, but due to the physics itself as it is seen to exist for simulations even in the orthogonal coordinate system.  By generating new grids with increased resolution in this region, the simulations were pushed to later time steps.   Further work will be undertaken to stabilise the simulations by modifying the neutral model to include extra diffusion in the poloidal plane.\\

\section{Conclusion}
The simulation of plasma behaviour in the divertor region is increasingly important as the power of tokamaks increases.  Melting needs to be avoided at all costs, so understanding and predicting the heat load is necessary.  A new coordinate system has been introduced that allows divertor geometry to be matched in simulation, while maintaining the benefits of a field-aligned coordinate system.  This coordinate system has been implemented in BOUT++ and tested for numerical accuracy as well as reproduction of reasonable physical behaviour.  These coordinates, in conjunction with a plasma and neutral turbulence model, will be used in future work to predict the heat load, radiated power, and evolution of detachment for devices such as ITER.
\\
\indent Though this work has focused on applications for tokamak physics, the definition of these coordinates is general in nature and depends only on choice of $\nu$ and $\eta$ to provide shifts to an orthogonal system.  It is conceivable to use the coordinates defined here to model any arbitrary 3D geometry where aligning coordinates to particular features of the physical system is desirable.

\subsubsection*{Acknowledgements}
We thank Maxim Umansky and Mikhail Dorf at Lawrence Livermore National Labs for useful suggestions in Cartesian gridding around the X-point geometry.  Some of the simulations were run using the ARCHER computing service through the Plasma HEC Consortium EPSRC grant number EP/L000237/1.   This work has received funding from the RCUK Energy Programme, grant number EP/I501045.  It has been carried out within the framework of the EUROfusion Consortium and has received funding from the Euratom research and training programme 2014-2018 under grant agreement No 633053.  The views and opinions expressed herein do not necessarily reflect those of the European Commission.\\

\bibliographystyle{unsrt} 
\bibliography{master}

\begin{thebibliography}{10}

\bibitem{Pitts2011}
R~A Pitts, S~Carpentier, F~Escourbiac, T~Hirai, V~Komarov, A~S Kukushkin,
  S~Lisgo, A~Loarte, M~Merola, R~Mitteau, A~R Raffray, M~Shimada, and P~C
  Stangeby.
\newblock {Physics basis and design of the ITER plasma-facing components}.
\newblock {\em Journal of Nuclear Materials}, pages 1--8, 2011.

\bibitem{Ricci2015}
Paolo Ricci.
\newblock {Simulation of the scrape-o ff layer region of tokamak devices}.
\newblock 81:1--14, 2015.

\bibitem{Dudson2009}
B.D. Dudson, M.V. Umansky, X.Q. Xu, P.B. Snyder, and H.R. Wilson.
\newblock {BOUT++: A framework for parallel plasma fluid simulations}.
\newblock {\em Computer Physics Communications}, 180(9):1467--1480, September
  2009.

\bibitem{Ribeiro2010}
Tiago~Tamissa Ribeiro and Bruce~D Scott.
\newblock {Conformal Tokamak Geometry for Turbulence Computations}.
\newblock 38(9):2159--2168, 2010.

\bibitem{Schneider1992}
R.~Schneider, D.~Reiter, H.P. Zehrfeld, B.~Braams, M.~Baelmans, J.~Geiger,
  H.~Kastelewicz, J.~Neuhauser, and R.~Wunderlich.
\newblock {jnurnalnf nuclear materials B2-EIRENE simulation of ASDEX and
  ASDEX-Upgrade scrape-off layer plasmas}.
\newblock {\em Journal of Nuclear Materials}, 198:810--815, 1992.

\bibitem{Radford1996}
G.~Radford, A.~Chankin, G.~Corrigan, R.~Simonini, J.~Spence, and A.~Taroni.
\newblock {The Particle and Heat Drift Fluxes and their Implementation into the
  EDGE2D Transport Code}.
\newblock {\em Contributions to Plasma Physics}, 36(2/3):187--191, 1996.

\bibitem{DHaeseleer1991}
W.D. D'haeseleer.
\newblock {\em Flux coordinates and magnetic field structure: a guide to a
  fundamental tool of plasma structure}.
\newblock Springer series in computational physics. Springer-Verlag, 1991.

\bibitem{Haeseleer1991}
W.~Haeseleer.
\newblock {\em Flux coordinates and magnetic field structure: a guide to a
  fundamental tool of plasma structure}.
\newblock Springer-Verlag, 1991.

\bibitem{Salari2000}
K.~Salari and P.~Knupp.
\newblock {Code Verification by the Method of Manufactured Solutions}.
\newblock {\em Sandia National Laboratories}, (Technical Report SAND2000-1444),
  2000.

\bibitem{Dudson2016}
B.~{Dudson}, J.~{Madsen}, J.~{Omotani}, P.~{Hill}, L.~{Easy}, and
  M.~{L{\o}iten}.
\newblock {Verification of BOUT++ by the Method of Manufactured Solutions}.
\newblock {\em ArXiv e-prints}, February 2016.

\bibitem{Lao1985}
L.L. Lao, H.~John, R.D. Stambaugh, a.G. Kellman, and W.~Pfeiffer.
\newblock {Reconstruction of current profile parameters and plasma shapes in
  tokamaks}.
\newblock {\em Nuclear Fusion}, 25(11):1611--1622, 1985.

\bibitem{Courant1967}
R.~Courant, K.~Friedrichs, and H.~Lewy.
\newblock {On the Partial Difference Equations of Mathematical Physics}.
\newblock {\em IBM Journal of Research and Development}, 11(2):215--234, 1967.

\bibitem{Yoshizawa2001}
A.~Yoshizawa, S.~Itoh, K.~Itoh, and N.~Yokoi.
\newblock {Turbulence theories and modelling of fluids and plasmas}.
\newblock {\em Plasma Physics and Controlled Fusion}, 43(3):R1--R144, March
  2001.

\bibitem{Leddy2015}
J.~Leddy, B.~Dudson, M.~Romanelli, and JET Contributors.
\newblock {On the validity of drift-reduced fluid models for tokamak plasma
  simulation}.
\newblock {\em Plasma Physics and Controlled Fusion}, 57(12):125016, 2015.

\bibitem{Simakov2004}
A.~N. Simakov and P.~J. Catto.
\newblock {Drift-ordered fluid equations for modelling collisional edge
  plasma}.
\newblock {\em Contributions to Plasma Physics}, 44(1):83--94, 2004.

\bibitem{Loizu2012}
J.~Loizu, P.~Ricci, F.~D. Halpern, and S.~Jolliet.
\newblock {Boundary conditions for plasma fluid models at the magnetic
  presheath entrance}.
\newblock {\em Physics of Plasmas}, 19(2012), 2012.

\bibitem{Loarte2001}
Alberto Loarte.
\newblock {Effects of divertor geometry on tokamak plasmas}.
\newblock {\em Plasma Physics and Controlled Fusion}, 43(6):R183--R224, 2001.

\bibitem{Stangeby2000}
P.~Stangeby.
\newblock {\em The plasma boundary of magnetic fusion devices}.
\newblock Institute of Physics Pub., 1 edition, 2000.

\bibitem{Togo2013}
S.~Togo, M.~Nakamura, Y.~Ogawa, and K.~Shimizu.
\newblock {Effects of Neutral Particles on the Stability of the Detachment
  Fronts in Divertor Plasmas}.
\newblock {\em The Japan Society of Plasma Science and Nuclear Fusion
  Research}, 8:1--4, 2013.

\bibitem{Lipschultz1995}
B~Lipschultz, J~Goetz, B~LaBombard, G~M McCracken, J~L Terry, M~Graf, R~S
  Granetz, D~Jablonski, C~Kurz, A~Niemczewski, and J~Snipes.
\newblock {Dissipative divertor operation in the Alcator C-Mod tokamak}.
\newblock {\em Journal of Nuclear Materials}, 220–222:50--61, 1995.

\end{thebibliography}

\end{document}